\documentclass[
 reprint,
superscriptaddress,
 amsmath,amssymb,
 aps,
 prx
]{revtex4-1}
\usepackage{color,soul}
\usepackage{float}
\usepackage{graphicx}
\usepackage[dvipsnames]{xcolor}
\usepackage{dcolumn}
\usepackage{bm}
\usepackage{soul}
\usepackage[framemethod=tikz]{mdframed}
\usepackage{tcolorbox}
\usepackage{tabularx}
\usepackage{array}
\usepackage{colortbl}

\tcbuselibrary{skins}
\newcommand{\specificthanks}[1]{\@fnsymbol{#1}}
\newcolumntype{Y}{>{\centering\arraybackslash}X}
\tcbset{tab2/.style={enhanced,fonttitle=\bfseries,fontupper=\normalsize\sffamily,
colback=white!90!gray,colframe=black!50!black,colbacktitle=black!70!white,
coltitle=white,center title}}

\usepackage{subcaption}
\usepackage{ragged2e}
\DeclareCaptionJustification{justified}{\justifying}
\begin{document}
\preprint{APS/123-QED}

\title{Effective compression of quantum braided circuits aided by ZX-Calculus}
\author{Michael Hanks}
\altaffiliation[]{Both authors contributed equally to this manuscript.\\ mpestarellas@nii.ac.jp}
\affiliation{National Institute of Informatics, 2-1-2 Hitotsubashi, Chiyoda-ku, Tokyo 101-8430, Japan}
\author{Marta P. Estarellas}
\altaffiliation[]{Both authors contributed equally to this manuscript.\\ mpestarellas@nii.ac.jp}
\affiliation{National Institute of Informatics, 2-1-2 Hitotsubashi, Chiyoda-ku, Tokyo 101-8430, Japan}
\author{William J. Munro}
\affiliation{NTT Basic Research Laboratories \& NTT Research Center for Theoretical Quantum Physics, NTT Corporation, 3-1 Morinosato-Wakamiya, Atsugi, Kanagawa 243-0198, Japan}
\affiliation{National Institute of Informatics, 2-1-2 Hitotsubashi, Chiyoda-ku, Tokyo 101-8430, Japan}
\author{Kae Nemoto}
\affiliation{National Institute of Informatics, 2-1-2 Hitotsubashi, Chiyoda-ku, Tokyo 101-8430, Japan}

\begin{abstract}
Mapping a quantum algorithm to any practical large-scale quantum computer will
require a sequence of compilations and optimizations. At the level of
fault-tolerant encoding, one likely requirement of this process is the
translation into a topological circuit, for which braided circuits represent
one candidate model. Given the large overhead associated with encoded circuits,
it is paramount to reduce their size in terms of computation time and qubit
number through circuit compression. While these optimizations have typically
been performed in the language of $3$-dimensional diagrams, such a
representation does not allow an efficient, general and scalable approach to
reduction or verification. We propose the use of the ZX-calculus as an
intermediate language for braided circuit compression, demonstrating advantage
by comparing results using this approach with those previously obtained for the
compression of $\vert A\rangle$ and $\vert Y\rangle$ state distillation
circuits. We then provide a benchmark of our method against a small set of
Clifford+T circuits, yielding compression percentages of $\sim77$\%. Our
results suggest that the overheads of braided, defect-based circuits are
comparable to those of their lattice-surgery counterparts, restoring the
potential of this model for surface code quantum computation.
\end{abstract}

\pacs{Valid PACS appear here}

\maketitle

\section{Introduction}
\label{sec:introduction}

\begin{figure}[ht!]
    \includegraphics[scale=0.5]{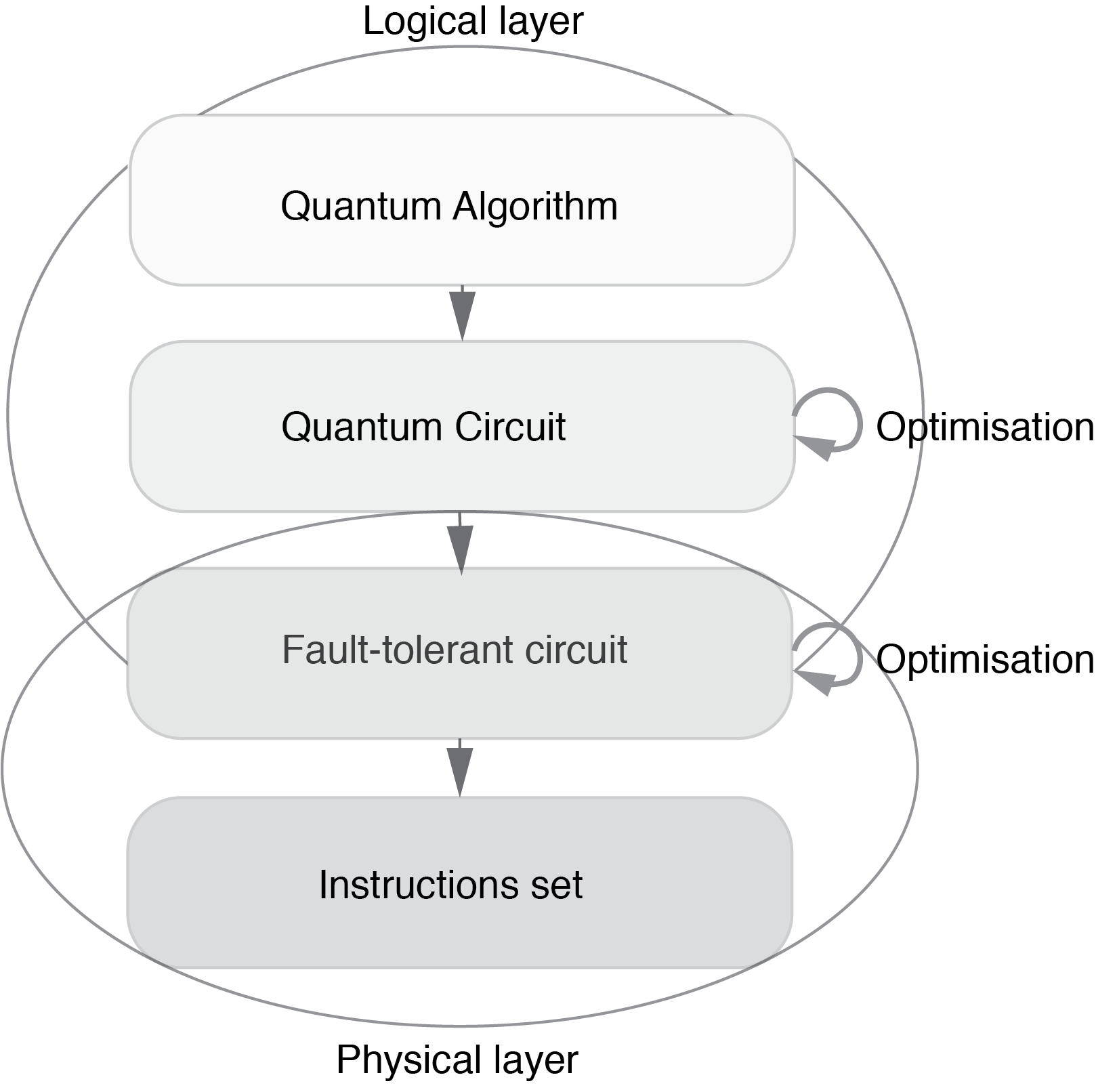}
    \caption{General compilation stack of a large-scale quantum computer
              architecture.}
    \label{fig:compilation}
\end{figure}

Quantum computers hold the promise of finding solutions to problems that cannot
be efficiently treated using the general classical model of computation
\cite{Deutsch85,Shor1997,grover1996}. Given the potential advantages this
technology has to offer, and its latest developments \cite{Arute2019}, global
efforts are currently focusing on the design of practical large-scale quantum
computer architectures that will allow for a real quantum advantage
\cite{Fowler12c}. On the software side, a common element to those designs is
the need to specify a quantum compilation process. As in classical computers, a
complete stack of consecutive translations is needed in order to map an
arbitrary quantum algorithm into a reduced set of machine operations (see
Figure~\ref{fig:compilation}). We identify that the general question of parsing
a quantum problem from a high-level description into machine language for
large-scale applications needs to involve at least three compilations. At the
highest level of abstraction, the quantum algorithm is written in a
human-readable language and the first compilation generates a quantum circuit
from a reduced set of universal quantum gates.  The second compilation is in
charge of making the computation robust to errors and thus requires the
inclusion of quantum error correction methods. At the lowest level, the
fault-tolerant circuit has to be translated into a sequence of
hardware-specific classical instructions (e.g. sequence of microwave pulses)
that leads the machine to effectively perform the desired computation.

A major challenge in the realization of a large-scale quantum computer is
related to the redundancy required by the error-correction codes and their need
of large numbers of physical qubits. One of the most promising error correcting
codes, offering accuracy threshold values of $\sim 1\%$, is the surface code
\cite{Bravyi98a,Freedman01a,Dennis02a}. Depending on the way the logical qubits
are encoded in the surface code the type of operations applied to the physical
qubits will differ leading to two main models of computation: braiding
(defect-based encoding) \cite{Fowler12f} and lattice surgery (planar-based
encoding) \cite{Horsman12a}. Each of these is represented as a different $3$D
diagram that keeps track of the operations (braiding, deformation and
merging/splitting) that need to be applied between the logical degrees of
freedom of each representation. Given that for large-scale purposes any
arbitrary optimized non-error corrected circuit will likely need to be
translated into a topological one, the optimization of such intermediate
circuit will be paramount in order to reduce its associated resources and
therefore relax the physical demands of the quantum device. 

Motivated by the search for low-overhead error-corrected circuits to bring
forward the regime of quantum practicality, we here focus on the compression of
defect-based topological quantum circuits. Given such a circuit, we want to
reduce the resources associated with the circuit volume, i.e. the time and
number of defects (and therefore, physical qubits). These optimizations have
typically been performed in the language of $3$-dimensional diagrams
\cite{Fowler12f,Paetznick13b}. However, due to the complicated nature of large
circuits expressed in this representation, it is difficult to intuit and keep
track of transformation rules. Importantly, this representation provides no
strategy to ensure the independence of errors arising from different local
gates in distillation circuits or to verify such circuits beyond Clifford-group
sub-segments. We here instead propose the use of the ZX-calculus representation
\cite{Coecke10a} as an alternative language for braided topological circuit
optimization. The ZX-calculus clarifies our intuition about topological
transformations and provides a straightforward method for the verification of
error independence and therefore fault-tolerance \cite{Chancellor16a}.
Importantly, we show how the primitive elements of the ZX-calculus can be
mapped directly to elements of a topological circuit each requiring specified
resources. Therefore using this route we can optimize the circuit with the
ZX-calculus and then convert back to the $3$D representation to apply final
arrangements and trivial transformations for a further packing of the braids.
We demonstrate how this approach allows the  compression of general braided
circuits, observing reductions of up to $\sim77$\%. We also demonstrate that
the resources required for these reduced braided structures are comparable to
those obtained for circuits based on lattice-surgery \cite{Litinski2019}.

Following an introduction to braided circuits
(Section~\ref{sec:braided_circuits}) and the ZX Calculus
(Section~\ref{sec:zx_calculus}), we demonstrate the use of the ZX Calculus as
an intermediate language for circuit reduction with two exemplar magic state
distillation circuits in Section~\ref{sec:exemplar_circuits}. We next present a
wider range of compression results and discuss the potential of hybrid braid
and lattice-surgery approaches in
Sections~\ref{sec:benchmarking_with_general_circuits}~and~\ref{sec:lattice_surgery}
respectively. Finally, in Section~\ref{sec:conclusions} we discuss the
implications of our results in the broader context of quantum circuit
compilation.

\section{Braided Circuits}
\label{sec:braided_circuits}

The surface code \cite{Bravyi98a,Freedman01a,Dennis02a} is one of the
most promising error correction codes for large-scale quantum applications. In
this code the physical qubits are arranged in a mesh-like $2D$ lattice and are
divided into two types: \textit{data qubits} and \textit{ancillae}.
{Data qubits} are identified with edges of the square lattice and the
{ancillae} are identified with either vertices or faces (sometimes named
\textit{plaquettes}). Parity measurements are repeatedly performed at the
{ancillae} to identify and track errors occurring in the
{data qubits}. The structure of the code is similar to the structure of
the letter H; on two sides (left and right) we have boundary vertices
with only three incident edges, while on the remaining two sides (top and
bottom) we have the remnants of face operators from which one boundary edge has
been removed. The former we call \textit{smooth} boundaries and the latter
\textit{rough}. Vertices and faces are stabilized by Pauli operators defined in
orthogonal bases respectively, such that they form two codes that together
protect from any arbitrary single-qubit error. Such a lattice is capable of
encoding one logical qubit fault-tolerantly, but to perform any
computation one needs to increase the logical degrees of freedom. Here we will
focus on the braided defect model as a surface-code based strategy for
fault-tolerant quantum computation.

\subsection{Defect Qubits}
\label{sub:defect_qubits-defect}

In the braided defect model additional logical degrees of freedom are
introduced by opening holes in the surface of the lattice such that the number
of logical qubits is increased by one. When we introduce multiple holes (or
defects) into the lattice, typically we do not make use of the entire code
space. Instead, we restrict the available logical operations so that we can
identify qubits locally with particular defect pairs. Local defect qubits can
be constructed with boundaries of either type (smooth or rough). We call
defects with one type of boundary \textit{primal} and the other \textit{dual},
and the bases of their respective logical operators are inverted.

Defects in the lattice of a surface code can be expanded by measuring
physical qubits across a region contiguous with the defect \cite{Fowler12c}, in
the basis of the stabiliser operations around the defect boundary. This will
identify string operators crossing the expanded region, and all such strings
are corrected outward to surround the new boundary.  Similarly, defects can be contracted by
initialising qubits and resuming stabiliser measurements in a region contiguous
with the defect boundary. The correction may then proceed as if all qubits in
the new region had been lost \cite{Stace09a,Stace10a,Barrett10a}. With the
ability to expand and contract, we are able to migrate defects about the
lattice. Such migration allows us to implement non-local \emph{braiding}
operations \cite{Briegel09a}. With braiding, string operators associated with
one defect pair can be made to \emph{wrap} around a pair with a different
boundary type (see blow up at the bottom of Figure~\ref{fig:3dCircuitExample}).
The effect of these two-qubit string operator correlations is to implement a
CNOT gate. Care must be taken to maintain the code distance during migration.

The mechanics of defect migration can also be applied to \emph{split}
and \emph{merge} defects, in a manner analogous to lattice surgery
\cite{Horsman12a}. Any string operator surrounding a defect, while a sub-region
of parity measurements is re-initialised across the diameter,
will necessarily surround both
resulting smaller defects. Similarly, any string operator in contact with the
boundary of a defect while such a sub-region is re-initialised will be in
contact with one of the two smaller defects. Conversely, when the migration of
two defects brings them together, the measurement of individual qubits between
the two defects will identify the parity of string operators crossing between
them. String operators in contact with one of two merging defects will of
course remain in contact with the larger composite region. In fact, the
extension of a defect itself can be viewed as a merge between an existing
logical qubit and a newly created logical qubit in a superposition state, while
a split has all the mechanics of contraction. These merge and split operations
are typically used for single-qubit initialisation and measurement, but can
also be applied between qubits. The direction of corrective operations from the
central sub-region between two defects will be a matter of convention.

\subsection{The 3D representation}
\label{sub:topological_quantum_circuits}

\begin{figure}[ht!]
\begin{center}
    \includegraphics[scale=0.25]{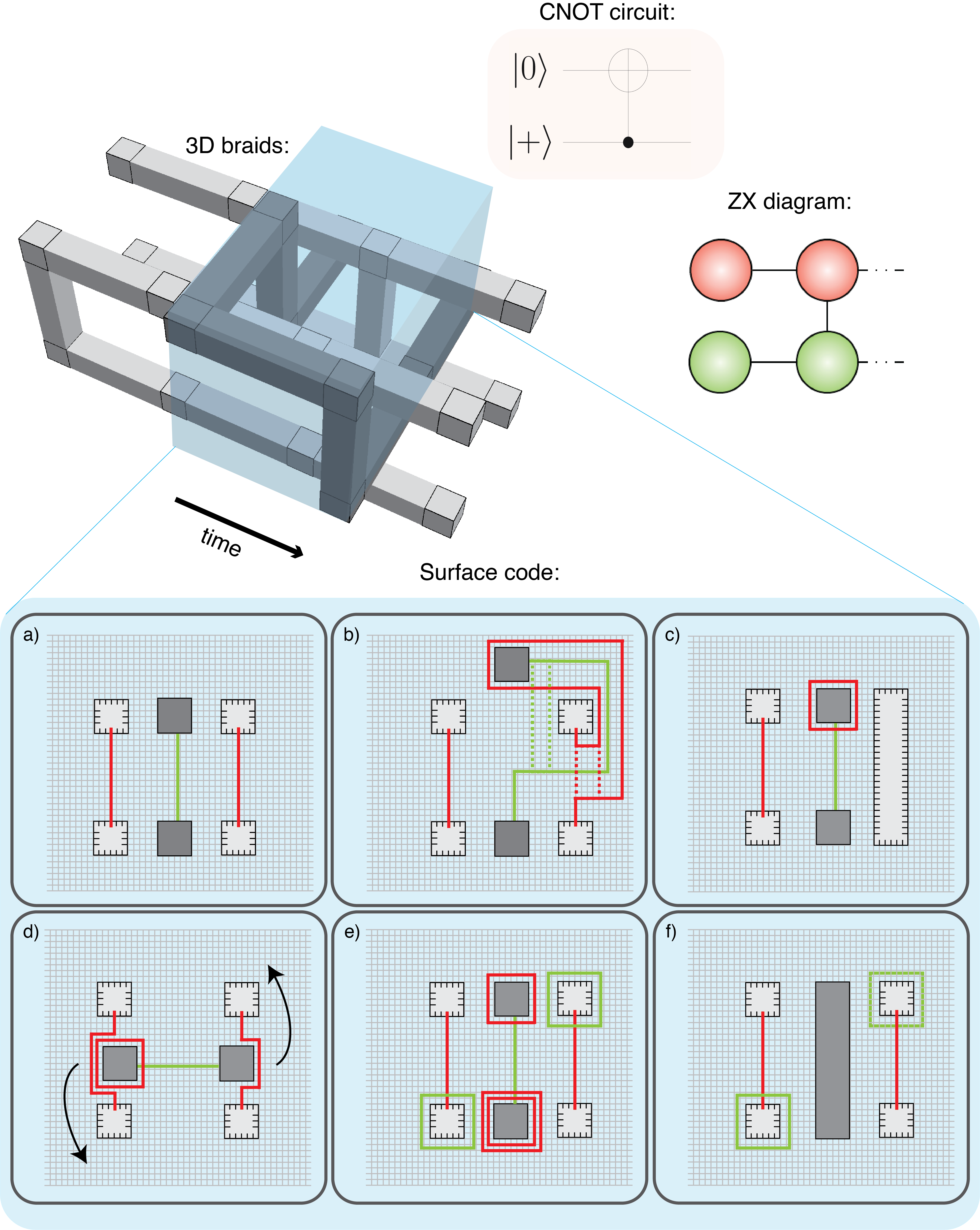}
    \caption{Example of a small quantum circuit performing a CNOT gate in the
    diagram, $3D$ and ZX-Calculus representations. Below, snapshots of the
    braiding process (note corrections are not shown): a) Two primal qubits
    initialised in the $\vert + \rangle$ and $\vert 0 \rangle$ states, and a
    dual qubit, initialised in the $\vert + \rangle$ state (the initial split
    operation for the dual qubit initialization is omitted) b) The dual defect
    braids with the first primal qubit. c) A merge operation between the two
    defects of the first primal qubit projects its state onto the dual qubit,
    and the merged defect is used to re-initialise the primal qubit in the
    $\vert 0 \rangle$ state. d, e) Braided interaction between the dual qubit
    and both primal qubits performs a CNOT gate with two targets. f) A merge
    operation between the two defects of the dual qubit measures it in the
    X-basis, so that the total operation corresponds to a CNOT from the first
    primal qubit to the second. }
    \label{fig:3dCircuitExample}
\end{center}
\end{figure}

3D diagrams are typically used to represent the behaviour of defects
opened in the surface code as they evolve across time. In such
representations, we ignore the individual physical qubits that make up the
surface code lattice, and represent the defect regions as \emph{pipes} of
appropriate relative width \cite{Briegel09a,Paetznick13b}. These pipes are
either light (primal) or dark (dual), and wind around one another in
three-dimensions such that a two-dimensional cross-section of the structure
displays the locations of the defect regions at a point in time. In
Figure~\ref{fig:3dCircuitExample} we show a braided
CNOT gate between two primal qubits (with a dual qubit as intermediary).

Verification of 3D structures is performed via `correlation surfaces'
\cite{Paler15a}, which track the evolution of logical operators across the
structure in time.  This method does not distinguish between circuits that
differ only by topology-preserving transformations.  Nor does it track the many
possible phase combinations arising from probabilistic teleported gates.  The
final state may be preserved, conditional on the outcomes of gate
teleportation, and it is possible to eliminate certain local operations
(including measurements), without noticing the effect on higher encoded logical
layers (as in distillation circuits).  We encounter precisely this issue when
comparing prior results for the Y- and A-state distillation circuits in
Section~\ref{sec:exemplar_circuits}: the former appears to have weakened the
protective structure of the distillation circuit, while the latter has been
limited at the positions of T-gate injections by a desire to conserve it.

No \emph{efficient} method of complete verification exists for
arbitrary quantum circuits. However, in the case of the 3D representation we
are faced with the further difficulty that the very definition of qubits can be
ambiguous. Looking ahead to Section~\ref{sub:Ydistillation}, we observe that
there are many ways to define qubits between defects in a 3D
structure.  When attempting to reconstruct a circuit-like representation, many
choices lead to counter-intuitive results --- such as the
measurement and destruction of `half' a qubit. These results can ultimately be
reconciled in terms of the equivalence between entanglement and the destruction
of logical degrees of freedom.  However, it remains difficult to interpret
reduced 3D structures as circuit diagrams and takes some significant effort
just to identify what is happening in the language of qubits and subcircuits,
let alone to verify the entire computation.

\subsection{Circuit Reduction}
\label{sub:circuit_reduction}

As a consequence of the tremendous progress in quantum technologies over the
last decade, efforts have now focused on the reduction of the resources
associated with large error-corrected quantum circuits. We may divide
topological circuit reduction into three categories: gate efficiency, static
reduction, and dynamic reduction. The first, gate efficiency, involves the
discovery of new and more efficient methods for implementing quantum gates on
encoded information. Examples of this first type are measurements based on
merging and splitting \cite{Fowler12c}, as opposed to isolating regions of the
lattice as if they were independent planar codes \cite{Fowler09a}. Other
examples are the direct implementation of the phase and Hadamard gates
\cite{Brown17a}, or the implementation of the CNOT gate, which may be performed
via braiding as mentioned above, but has also been performed via
teleportation-inspired `junctions' \cite{Paler12a}. The second type,
\emph{static} reduction, is circuit reduction in its usual form; given a set of
gates and transformation rules or identities, we seek to re-arrange the
elements of the circuit so that it occupies a smaller resource--time volume.
Examples are the direct `compaction' of defect pipes \cite{Paetznick13b}, the
use of `bridging' to eliminate redundant portions of the circuit
\cite{Fowler12f}, or the selective arrangement of the components of magic state
distilleries \cite{Ding18a}. The dynamic element of circuit reduction, in
contrast to the static element, seeks less to reduce any single circuit and
more to find an approach that can be applied to any circuit to reduce the
overhead costs associated with information received at run-time. One example is
the use of online-scheduling \cite{Paler17b}, while another is the use of
selective-route teleportation \cite{Fowler12e} to fix the time-cost of a
circuit with probabilistic elements.
  
One approach that has been used to simplify the reduction (and partial
verification) of topological quantum circuits is the `ICM' representation
\cite{Krivine15a}. This representation first divides the circuit into the three
distinct phases of `initialisation', `CNOT gates', and `measurement'. Once this
has been achieved, and making use of selective-route teleportation to fix the
form of the circuit, the CNOT-restricted segment of the circuit can be reduced
and verified with `stabiliser' \cite{Paler18b} and `correlation-surface'
methods \cite{Paler15a}. 

We here restrict ourselves to the static reduction of general quantum circuits;
we will not make use of the ICM representation or related techniques on
restricted circuit classes. As in previous works, we assess the circuit
compression in terms of reducing the volume of an imaginary bounding box that
surrounds the 3D structure. This volume, in units of the code distance `d', has
a direct correspondence with both the number of necessary physical qubits and
the total computation time.

\section{The ZX-Calculus}
\label{sec:zx_calculus}

The ZX-Calculus is a tensor-network-based diagrammatic language introduced by
Coecke and Duncan \cite{Coecke08a,Coecke11a} that allows for a high level and
intuitive representation of pure state qubit quantum processes. This
representation is universal and complete for Clifford+T quantum mechanics
\cite{Backens14b,Coecke18a}, of direct concern to surface-code quantum
computing, and provides a complete set of rewrite rules for reasoning about
equivalent diagrams. The application of the ZX-Calculus has proven fruitful in
the representation of surface codes with lattice surgery \cite{deBeaudrap17a},
for design and analysis of topological quantum algorithms \cite{Horsman11a},
and as a method for reducing the number of T-gates in a quantum circuit
\cite{kissinger2019tcount}.

\begin{figure}[ht!]
    \centering
    \includegraphics[scale=0.6]{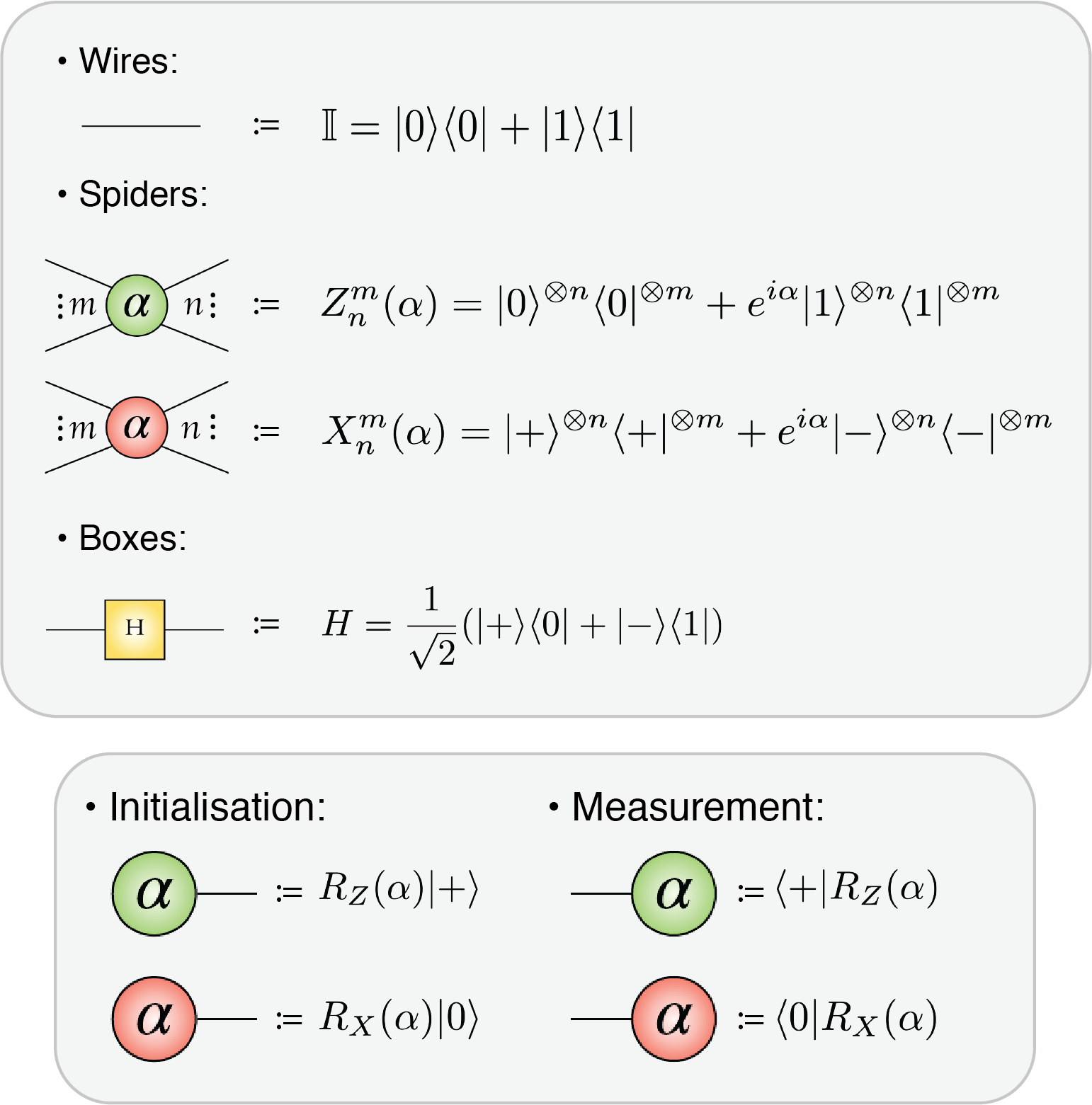}
    \caption{ZX-Calculus elements and their associated linear operations.
    Initialisation and measurement operations in the Z and X basis.}
    \label{fig:ZX_elements}
\end{figure}

By convention, ZX diagrams are read timewise from left (inputs) to right
(outputs). They are connected, undirected graphs, with edges representing
single qubits and labelled green (red) nodes representing transformations and
measurements of these qubits in the Z (X) basis (see
Figure~\ref{fig:ZX_elements}). Nodes may have many edges and are referred to as
\textit{spiders}, and additionally a yellow box is used as short-hand to
represent the single-qubit Hadamard gate.  Placing elements above and below one
another denotes their tensor product ($\otimes$), while connecting nodes
denotes their composition ($\circ$). The circuit is preserved under any
bending, stretching or twisting of wires; nodes may be rearranged as long as
the elements and connectivity of the diagram are conserved (graph isomorphism).
Qubit initialisation and projective measurement are common operations
represented as shown in Figure~\ref{fig:ZX_elements}.

\begin{figure*}[ht!]
  \centering
    \includegraphics[scale=0.6]{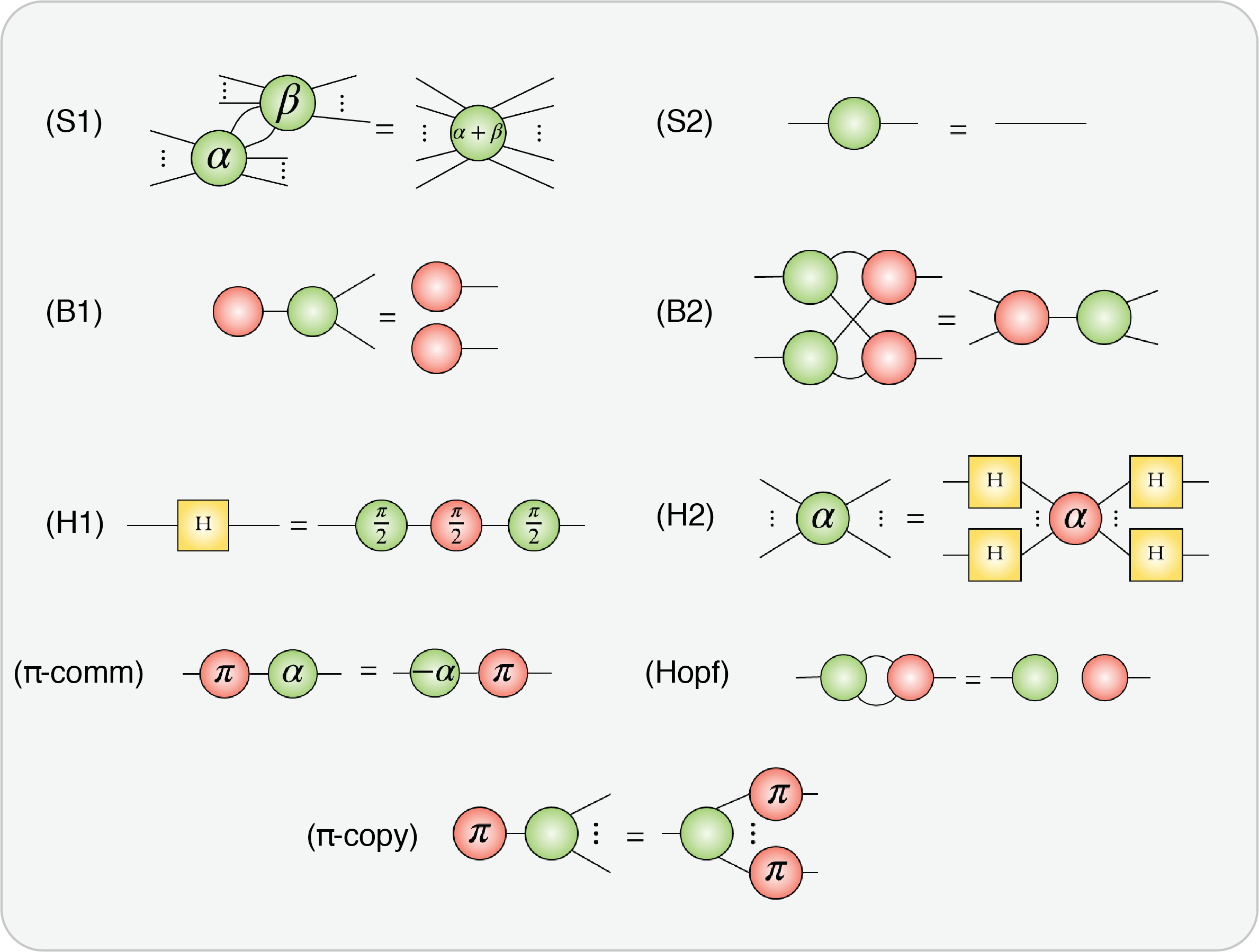}
    \caption{ZX-Calculus rules. This set of reversible axioms or rewrite rules is complete for Clifford circuits.}
    \label{fig:ZX_axioms}
\end{figure*}

In Figure~\ref{fig:ZX_axioms} we outline the axioms needed for Clifford
completeness. Each of these rewrite rules holds under the exchange of red and
green nodes or when the diagrams are reflected. The original axioms include
disconnected components (``scalars'') to normalize equivalent structures
\cite{Backens15a}. These scalars do not affect the connectivity of the diagram,
and we may therefore ignore them in our resource estimates. However, it should
be remembered that probabilistic elements and associated corrections remain
present.

Our approach to compression relies on the translation of our circuit into this
graph representation. The initial translation is straightforward when the
circuit is expressed in Clifford+T form. Following this, the
reduction of a ZX-Calculus graph consists of the application of a sequence of rewrite
rules (see Fig.~\ref{fig:ZX_axioms}) to minimise the number of red, green and
yellow nodes. When dealing with braided
circuits, this minimization has two additional constraints: 1) we want to
minimize the number of $\pi/4$ and $\pi/8$ gates (S and T gates, respectively),
as their presence has overhead associated with distilleries
(described in Section~\ref{sec:exemplar_circuits}); 2) we want to
minimize the number of Hadamard gates, also due to their additional
overheads. The ZX-graph reduction of small circuits, such as
those we present in Section~\ref{sec:exemplar_circuits}, can be easily done
by hand. In constrast, for larger circuits such as those in
Section~\ref{sec:benchmarking_with_general_circuits} one needs to use computer aided tools.
We found it helpful to put several of the larger circuits through an initial
round of automated reduction with PYZX, an open source python library designed
to reduce, validate and visualize ZX-Calculus diagrams \cite{pyzxgit}. PYZX
applies a recursive, greedy algorithm \cite{kissinger2019}. Though the strategy
of the PYZX library achieves significant reductions, it does not necessarily
take into account the additional gate costs mentioned above (for instance,
the reduced graphs of PYZX tend to have many Hadamard gates).  Nonetheless,
having reduced the overall graph size, it became feasible in isolation to
tackle the $\pi/4$, $\pi/8$ and Hadamard gates by hand.

\subsection{Translating from the ZX Calculus to the 3D Topological Representation}
\label{sub:translating_between_zx_and_3d_topological_representations}

\begin{figure}[ht!]
    \centering
    \includegraphics[scale=0.5]{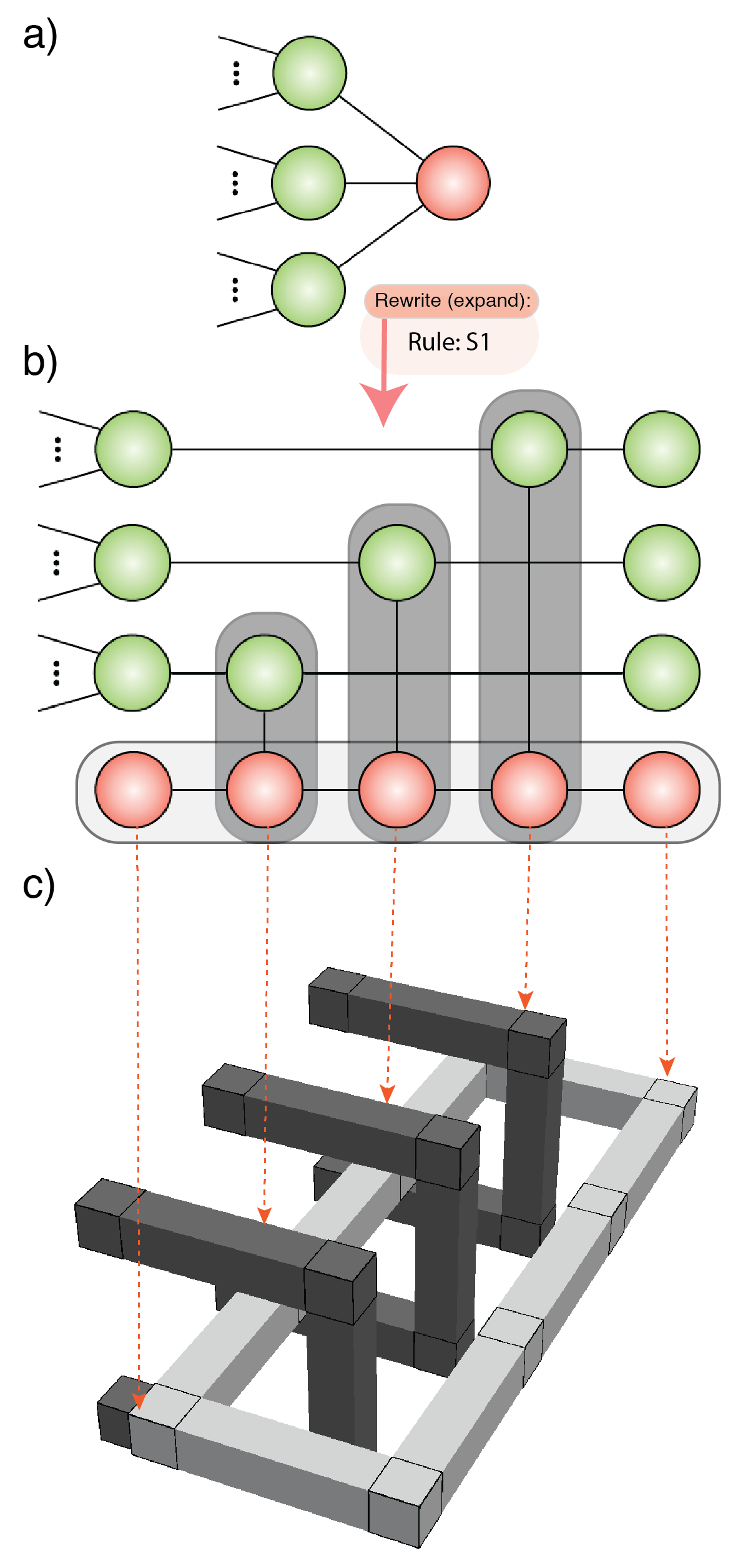}
    \caption{Translation example. a) Partial ZX circuit illustrating a typical
    connected red spider node. b) Expansion of the same circuit through the
    application of the S1 rule leading to a set of CNOT gates, initialisations,
    and measurements. CNOT gate identification allows us to reduce the number
    of merge and split operations (and associated qubits) through braiding. c)
    Associated 3D encoded circuit where the primal defect is identified with
    the red spider node. The identification of spider nodes with closed loops
    allows the \textit{nodes} of the ZX graph to be interpreted as qubits (in
    lieu of edges).}
    \label{fig:ZX_translationDemo}
\end{figure}

Having obtained a reduced ZX-calculus graph, we are faced with the task of
translating this result back into a 3D topological encoded quantum circuit. In
contrast to planar lattice surgery \cite{deBeaudrap17a}, the problem of
correspondence between elements presents a problem unique to defect qubits: the
basis of a merge or split (present in red and green spiders) is determined by
the type of defect (primal or dual), so that some conversion must take place
between adjacent ZX nodes in distinct bases. To overcome this challenge we
propose a new interpretation of the ZX graph: for braided logic, we view the
nodes of a ZX graph as qubits (with red nodes as primal defect pairs and green
nodes as dual defect pairs), edges between nodes of distinct colours
as braided interactions between these qubits (see
Figure~\ref{fig:ZX_translationDemo}), and edges between nodes of the same
colour as junctions \cite{Raussendorf07b}. Our new interpretation is possible
only because the ZX-calculus assumes positive-parity measurements at
merge-points (or applies corrections), such that parities propagate onto the
output. Note that, due to constraints on inter-defect distances, the number of
edges still forms a lower bound on the surface area enclosed by defect loops. 
We do suspect however that the computational cost associated with problem of
defect arrangement should be reduced along with the number of qubits.

Considering all nodes of degree one or two as local operations or state
injections, several axioms (or rules) of the ZX-calculus (see Figure \ref{fig:ZX_axioms}) now take on familiar forms as
transformations of topologically encoded circuits. For example, the $\pi$-copy rule describes the propagation of correlation surfaces \cite{Paler15a}, while
the Hopf rule for edge pairs between adjacent nodes becomes the parity of the
winding number in the braided circuit. The S2 rule
describes the triviality of a pair of operations swapping back and forth
between primal and dual qubits. The B2 rule is non-trivial, though we find the
structure of this element in Figure~$12$ of \cite{Raussendorf07b}.

\section{Exemplar Circuits}
\label{sec:exemplar_circuits}

We are now equipped to apply the compression method described in
Section~\ref{sec:zx_calculus} to some relevant exemplary circuits. For this we
have chosen two important elements of many large quantum circuits: Y-state
preparation for the application of the $\pi/4$ gate and A-state distillation
for the application of the $\pi/8$ gate. Though the search for low-overhead fractional phase gates is ongoing
\cite{Meier12a,Bravyi12a,Jones13a,Jones13b,Jones13c,Eastin13a,Paetznick13a}, we
focus here on the simple cases where the $15$-qubit Reed-Muller code is used
for A-state distillation and Steane's code for the Y-state \footnote{It is
worth emphasising that these circuits do not represent the state-of-the-art in
terms of overall cost; we select them here primarily for the purpose of
comparative reduction.}. We note that recent work has proposed a
$\pi/4$ gate implementation to circumvent the need for distillation
\cite{Brown17a}. These circuits have not only been chosen due to the
critical importance of these states in surface code quantum computation but also because they allow us to compare our method to the one presented in \cite{Fowler12c}.

The surface code allows the $\sigma_{X}$, $\sigma_{Z}$, Hadamard, and CNOT
gates transversally. To achieve a universal quantum gate set we may add the
$\pi/8$ gate. The current method of achieving this gate was proposed by Bravyi
and Kitaev in $2005$ \cite{Bravyi05a} and makes use of gate teleportation
\cite{Bennett93a,Gottesman99a,Gottesman99b}. The key is that the ancillary
state containing information about the gate is a \textit{known} state, the
A-state. This means we can use an error detection protocol to improve its
fidelity to levels consistent with other logical qubits before applying its
effects via teleportation. Such ancillary states are referred to as
\textit{magic states}. Importantly, Bravyi and Kitaev showed that only Clifford
gates $\{H,\pi/4,CNOT\}$ were required in the distillation protocol. Further,
while $\pi/4$ gates are required for A-state distillation, only $\sigma_{z}$
gates are required to play the equivalent role for $\pi/4$ gates in Y-state
distillation.

The teleported gates are achieved in 3D topological circuits via \textit{state
injection}. This procedure measures the two-qubit $\hat{Z}_{1}\hat{Z}_{2}$
operator through a merge operation. On a parity result of zero the phase is
inherited by the target, while on a parity result of one a corrective phase
rotation with double the target angle is required. To perform the projective
$\hat{Z}_{1}\hat{Z}_{2}$ measurement in a 3D topological quantum circuit, the
single qubit on which a faulty $\pi/4$ or $\pi/8$ gate is initially applied
must be `grown' \cite{Fowler12c, Fowler09a, Dennis02a, Fowler2009} into a
defect pair so that a merge operation with the target can be performed. This
growth is represented diagrammatically as a pair of pyramidal structures
emerging in opposite directions from a point. In this paper we use the
convention that such pyramidal structures are coloured red for $\pi/8$ gates,
and green for $\pi/4$ gates. The $\pi/8$ gates may require trailing $\pi/4$
corrections. Space has been allowed for these corrections in the circuits to
follow, though we do not show them explicitly.

\subsection{$\vert Y \rangle$ distillation circuit}
\label{sub:Ydistillation}

\begin{figure*}[ht!]
    \centering
    \includegraphics[scale=0.23]{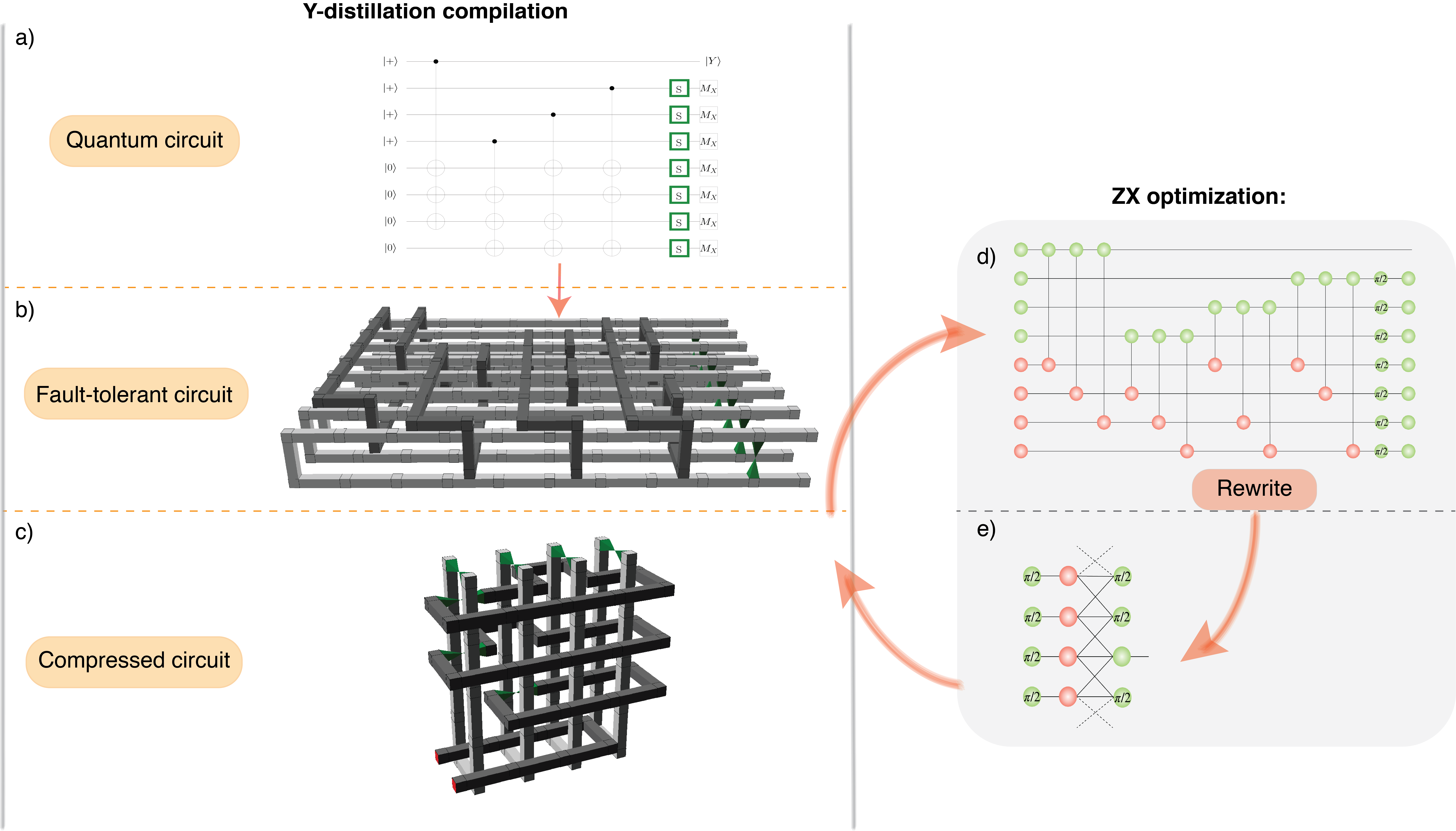}
    \caption{Compilation and optimization of the Y distillation circuit: a)
    Initial circuit in the diagram form. b) Initial circuit in the braided
    representation. c) Reduced circuit in the braided representation after
    applying our compression method. d) Initial circuit in the ZX-Calculus
    representation. e) Reduced circuit in the ZX graph representation. Note
    that the dashed lines indicate periodic boundaries.}
    \label{fig:Fig6}
\end{figure*}

In surface code quantum computing, the Y-state, $\sqrt{2}\vert Y \rangle =
\vert 0 \rangle + e^{i\pi/2} \vert 1 \rangle$, is used to implement the phase
($\pi/4$) gate via gate teleportation. Even though this gate does not form part
of the universal set of gates, a possible phase-gate correction is required for
every application of the $\pi/8$ gate. Some subcircuit for Y-state preparation is
therefore expected to be ubiquitous in quantum computing, and its reduction is
desirable. The preparation of the high-fidelity Y-state may be achieved either
directly \cite{Brown17a} or via one
of a series of distillation codes. For simplicity and in aid of our explanation, here we focus on the most
straightforward distillation approach, the direct application of the Steane quantum error
correction code in a new logical layer (see Figure~\ref{fig:Fig6}-a).

Mapping the Y-state distillation circuit to the 3D representation is
straightforward, as it only involves CNOT and phase gates (see
Figure~\ref{fig:Fig6}-b). CNOT gates are represented as dual defects braiding
the target primal qubits, while phase gates are performed through state
injection, as described above. Similarly, we can represent the original circuit
in the ZX language (see Figure~\ref{fig:Fig6}-d). CNOT gates are then
represented as a connected pair of green (control) and red (target) nodes
between appropriate qubit wires, while phase gates are represented as green
nodes with phase labels.

Through application of rules S1 and S2, we are able to reduce the ZX diagram
for the Y-state distillation circuit to that of Figure~\ref{fig:Fig6}-e. It is
important not to break the protection of the Steane code by over-optimizing the
ZX diagram. To ensure the code structure is maintained, we can validate the
code through the application of the $\pi$-copy rule. This rule allows us to
easily visualize how $\pi$ errors propagate forward through the diagram in time
\cite{Chancellor16a} to check whether the code still provides a unique syndrome
for each correctable error. Once the ZX diagram is reduced, and aided by the
interpretation introduced in
Section~\ref{sub:translating_between_zx_and_3d_topological_representations}, we
are ready to translate the reduced $\vert Y\rangle$ distillation circuit back
to the braided representation.

The ZX diagram does not provide any information about the time/space direction
of the sub-diagram that sits between the inputs and outputs. Therefore, nodes
can be arranged anywhere in a 2D plane so long as the connectivity is
preserved. When translating to the 3D representation, this means we must decide
how to arrange the braided components in space and time, selecting a causal
structure. In general, these decisions will depend on the relative costs
associated with time and physical qubit number.

Translating the reduced Y-state distillation circuit, we obtain the 3D
topological circuit shown in Figure~\ref{fig:Fig6}-c. The dimensions of this
circuit in units of the code distance, \textit{d}, are $2\times4\times4$, for a
total volume of $32$. This compares with a direct translation from the circuit
diagram (Figure~\ref{fig:Fig6}-a), for which we obtain a circuit of dimensions
$1.5\times8\times9$, for a total volume of $108$. We note that, though the
circuit is too small to give a sense of the general trade-off between qubits
and time, we are still free to re-orient the circuit to emphasise one or the
other resource. In this case, simple re-orientation allows either a
$4$-timestep, $8$-qubit circuit or a $2$-timestep, $16$-qubit circuit.

The Y-state distillation circuit is a natural initial test for any reduction
procedure, and in 2012 Fowler and Devitt \cite{Fowler12f} addressed it to
propose the 3D transformation rule `bridging'. The size of the circuit thus
obtained was $2\times1.5\times3$ for a volume of $9$. Clearly this is smaller
than the $32$-volume circuit we were able to obtain, and we now address the
sources of this discrepancy.

The first of the differences between the two methods arises in the encoding of
qubits in 3D space; braided circuits typically encode qubits in local defect
pairs, forming a subsystem code and ignoring additional `gauge' degrees of
freedom. This encoding decision is enforced for all transformations from a
ZX-graph to a 3D circuit, just as it would be for any other intermediate
representation. The 3D bridging transformation proposed by Fowler and Devitt,
however, breaks this restriction, and is therefore able to encode a greater
amount of information in the same space. It remains an interesting open
question what restrictions apply to the use of these additional degrees of
freedom in translation from the ZX-language, and we intend to investigate this
point further in a forthcoming paper.

The second difference between the result of \cite{Fowler12f} and our
own is that we have placed limits on the circuit reduction that were not
present in their work. These limits maintain the independence of error
syndromes for the Steane code, and preserve the basis of the output state.
Following a set of single-qubit operators across the result of
\cite{Fowler12f}, we draw two important conclusions: firstly, errors
from injected gates no longer result in independent syndromes, so that the
order of detectable operations and the logical gate fidelity of the Steane code
are reduced; secondly, the final state is in a basis orthogonal to the $\vert
Y\rangle$ state --- a consequence of swapping the basis of the code --- so that
a trailing Hadamard gate and additional volume will be necessary. This
difference highlights the importance of the procedure provided by the
ZX-calculus for verifying error independence, and we expect this to become more
relevant as circuit sizes increase.

\subsection{$\vert A \rangle$ distillation circuit}
\label{sub:Adistillation}

\begin{figure*}[ht!]
    \centering
    \includegraphics[scale=0.25]{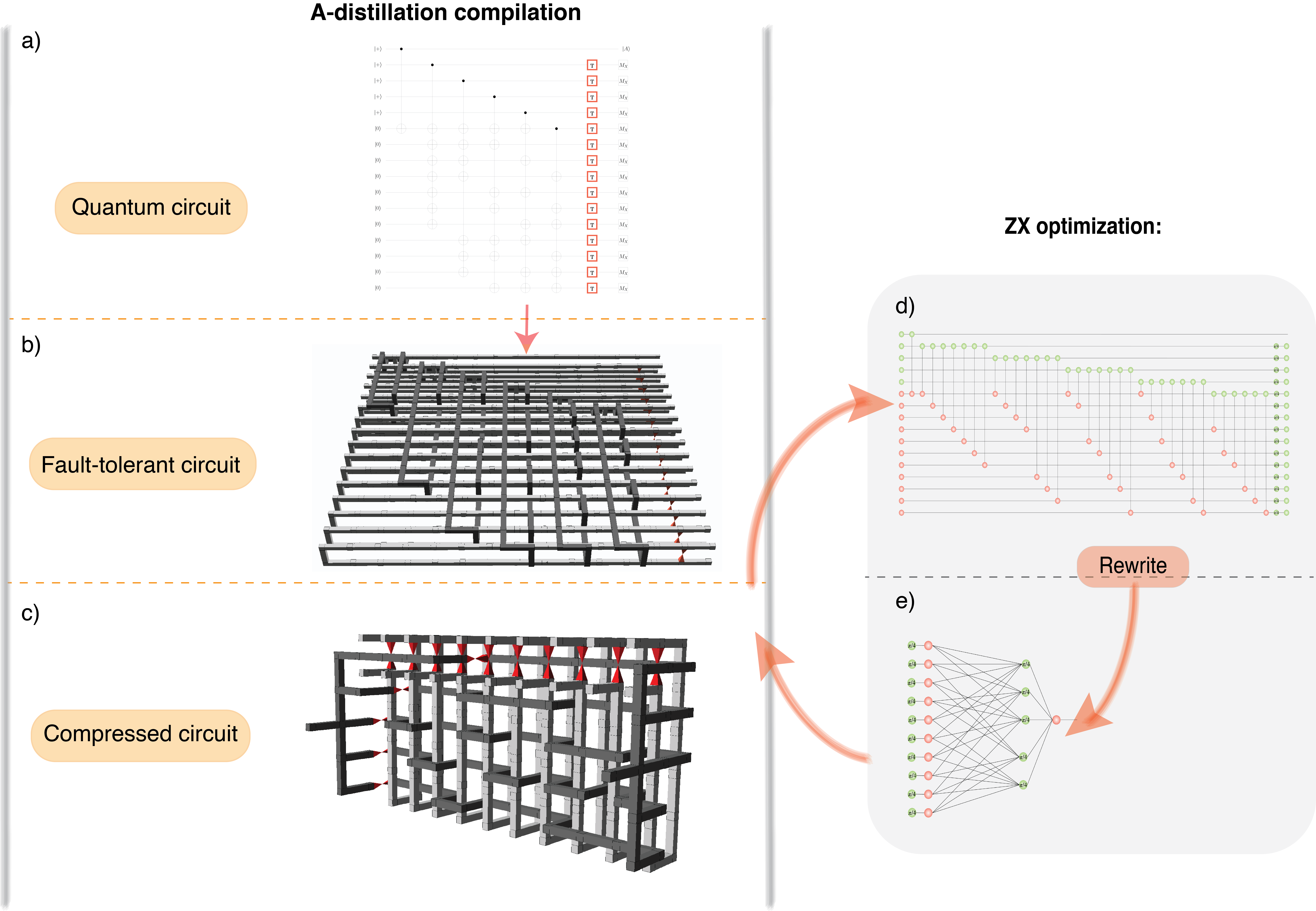}
    \caption{Compilation and optimization of the A distillation circuit: a)
    Initial circuit in the diagram form. b) Initial circuit in the braided
    representation. c) Reduced circuit in the braided representation after
    applying our compression method. d) Initial circuit in the ZX-Calculus
    representation. e) Reduced circuit in the ZX graph representation. Note
    that the dashed lines indicate periodic boundaries.}
    \label{fig:AdistCompilation}
\end{figure*}

Having established that the ZX-Calculus can be used to reduce and verify the
Y-state distillation sub-circuit, let us now turn to the more
expensive problem of A-state distillation for the $\pi/8$ gate. The A-state,
$\sqrt{2}\vert A \rangle = \vert 0 \rangle + e^{i\pi/4} \vert 1 \rangle$, is
used to implement the $\pi/8$ gate, the most common final element of a
universal quantum gate set incorporating also the Clifford gates (CNOT, X, H,
and S). Once again, there are a number of distillation codes that could be
chosen, and for simplicity we focus on the most conceptually simple: the direct
application of the 15-qubit quantum Reed--Muller code, depicted in
Figure~\ref{fig:AdistCompilation}-a.

The ZX-calculus axioms shown in Figure~\ref{fig:ZX_axioms} are complete only
for the Clifford gates. To achieve a complete set of axioms for the Clifford+T
gate set, several additions must be made \cite{Backens14b,Coecke18a}. While it
would be possible to identify structural identities corresponding to these
additional axioms, their complexity makes their manual discovery and
application difficult. Fortunately for our example, we are forearmed with
knowledge of the quantum Reed--Muller code structure and we know that it will
not be possible to reduce the number of T-gates without breaking the structure
of the code, and so in this case we need not attempt to apply the remaining
axioms. Once again, through application of S1 and S2, we have been able to
considerably reduce the complexity of the ZX diagram for the $\vert A \rangle$
distillation circuit as shown in Figure~\ref{fig:AdistCompilation}-e.
After translating the ZX-reduced A-state distillation circuit back
into the 3D representation, we obtain a reduced circuit with a total volume of
$125$. From an initial volume of $360$, the compression of this circuit yields
an outstanding compression rate of $65$\%.

As for the Y-state in Section~\ref{sub:Ydistillation}, a reduced 3D structure
for the A-state distillation circuit was given in \cite{Fowler12f}. For this
circuit, however, the reduction that was achieved with the topology preserving
transformations in 3D space was more limited, achieving a reduced volume of
$15.5\times 5.5\times 2.5 = 213.125$, or a $32$\% reduction. In stark contrast
to the previous example of Y-state distillation, the $125$-volume circuit we
obtain for A-state distillation is a further $41$\% smaller than the previously
reduced circuit. We note that the striking difference between these two
compression rates highlights the capabilities of the ZX-Calculus as an
intermediate language for braided circuit optimization. The limited reduction
attained when the optimization had been performed in the 3D
representation was a result of the presence of $\pi/8$ gates, which served to
partition the circuit into isolated regions, limiting applicable structural
transformations.


\section{Benchmarking with General Circuits}
\label{sec:benchmarking_with_general_circuits}

\begin{figure*}
    \centering
    \includegraphics[scale=0.80]{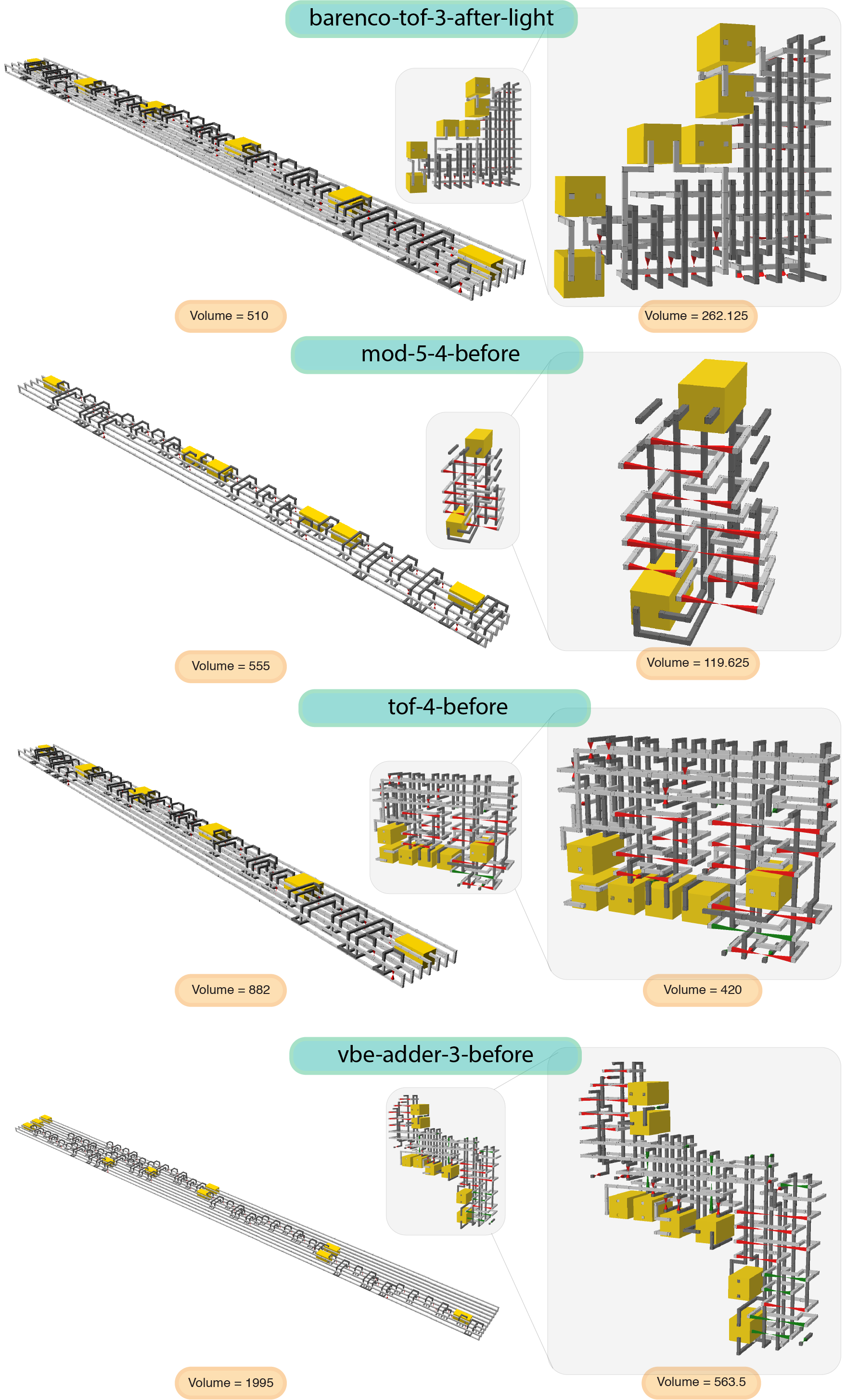}
    \caption{3D structures of the benchmarked circuits before and after
    compression and their respective volumes. We show the scaled structure to
    be compared with the initial non-reduced circuit and then a blow-up for
    detail.}
    \label{fig:benchmarks}
\end{figure*}

So far we have seen the application of the ZX-Calculus to circuit-reduction and
verification in two examples: the Y-state and A-state distillation circuits. To
show the generality of our approach, we have selected a small set of arbitrary
Clifford+T circuits as benchmarks to assess the performance of our circuit
optimization method. In Table~\ref{benchmarks}, in addition to the distillation
circuits, we outline four circuits with their respective original volumes in
the 3D representation, ($\text{Vol}_{init}$), and the reduced volume after
application of ZX-Calculus, ($\text{Vol}_{opt}$). Our results show compression
percentages of up to $\sim77\%$, giving a very promising indication of the
viability of our method for braided circuit compression against other
approaches. The corresponding 3D structures of such circuits before and after
optimization are shown in Fig. \ref{fig:benchmarks}. To account for
fault-tolerance, the volume-costs of magic state distillation for T-gates
should be added to the total volume of these additional circuits. In such
cases, that sum would represent an upper bound, as the reduction of a larger ZX
graph including the distillation circuits for the S and T gates could
potentially yield a smaller structure than the simple sum of the separate
volumes. 

\begin{table}[ht!]
  \begin{tcolorbox}[tab2,tabularx*={\renewcommand{\arraystretch}{1.5}}{p{10.5em}|p{5em}|X},title={Benchmarks},boxrule=0.9pt]
    \bf{Circuit} & \bf{$\text{Vol}_{init}$} & \bf{$\text{Vol}_{opt}$} \\\hline
    Y-distillation & 108 & 32 (-70.3\%) \\\hline
    A-distillation & 360 & 125 (-65\%) \\\hline
    barenco-tof-3-after-light & 510 & 262.125 (-48.6\%) \\\hline
    mod-5-4-before & 555 & 119.625 (-77.4\%) \\\hline
    tof-4-before & 882 & 420 (-52.4\%) \\\hline
    vbe-adder-3-before & 1995 & 563.5 (-72\%) \\\hline
  \end{tcolorbox}
  \caption{Volumes before and after optimization for a set of circuits from the
    \textit{PYZX} circuit database \cite{pyzxgit}.}
  \label{benchmarks}
\end{table}

\section{Further packing and lattice-surgery}
\label{sec:lattice_surgery}

While most of our compression is attained at the level of reduction of the ZX
diagram, when translating back to the 3D representation we have some freedom in
the way we arrange the braids in space. This arrangement depends on their
connectivity, so that an optimal packing of the structure allows us to further
reduce the volume. In addition, we can reinterpret some parts of our reduced
ZX-graph as \textit{merge} and \textit{split} operations between defects
analogous to lattice-surgery. Whether it is advantageous for volume reduction
to interpret a node as a braid loop or as such a defect-surgery operation
depends on the arrangement of the surrounding geometric structure. Note that
the ZX spider rule (S1) allows us to divide a single node into `braid' and
`surgery' regions. The surgery approach was in fact applied in the compression
of the $\vert A\rangle$ distillation circuit from
Fig.~\ref{fig:AdistCompilation}, and is implicit in our decision to treat
degree-one and degree-two nodes as local operations or state injections.

\begin{figure}[ht!]
    \centering
    \includegraphics[scale=0.5]{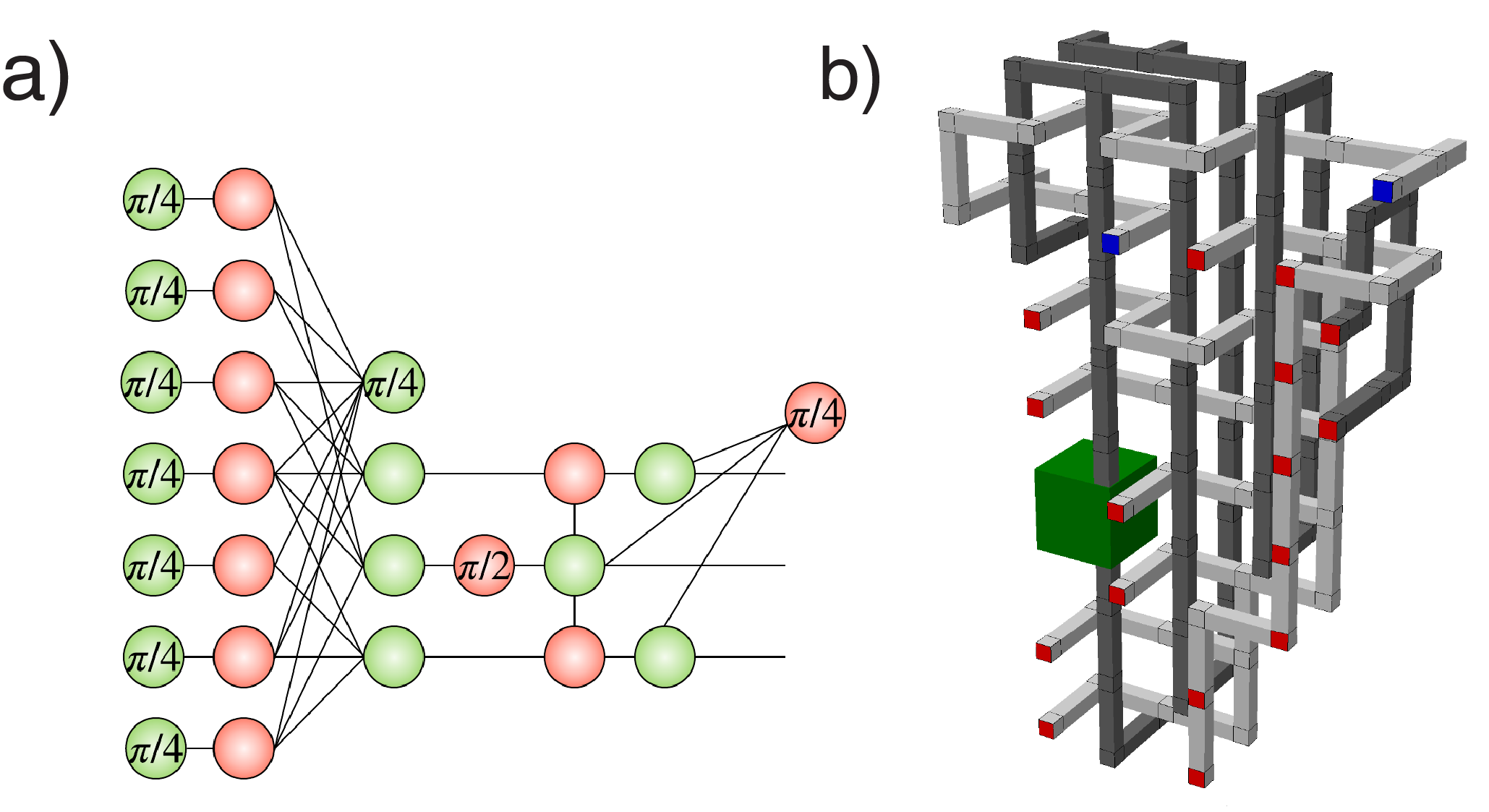}
    \caption{a) Reduced ZX diagram for the $\vert CCZ\rangle$-catalyzed $2\vert
    T\rangle$ factory from \cite{gidney2019efficient}. b) Corresponding 3D
    structure after packing and applying the hybridized approach for braiding
    and defect-surgery discussed in Section~\ref{sec:lattice_surgery}. Here
    we follow Figure~13 of \cite{gidney2019efficient} in substituting a
    green box for a single-qubit Clifford gate. We have marked the inputs of T-states in red, and the recycled T-state for catalysis in blue.}
    \label{fig:CCZfactory}
\end{figure}

To date, the most efficient T-gate distillation circuit is the $\vert
CCZ\rangle$-catalyzed $2\vert T\rangle$ lattice-surgery based circuit
depicted in Figure~18 of
\cite{gidney2019efficient}, with a volume of $120$
\footnote{In Figure~1 of \cite{gidney2019efficient}, the authors suggest outer bounding
boxes for the CCZ factory and catalyzed 2-T circuit of
$3\times 6\times 5.5 = 99$ and $4\times 4\times 6.5 = 104$ respectively, for a combined volume of
$203$.  However, it is possible that some space may be saved when these
components are packed or tiled within a larger circuit.  To account for this,
we use a highly conservative estimate for the volume, derived from Figure~18 of
\cite{gidney2019efficient}: we assign the CCZ factory a volume of
$2\times 5\times 6 = 60$ and the catalyzed 2-T circuit a volume of
$3\times 4\times 5=60$, for a combined total of $120$.}.
After we apply our
compression method to the same initial circuit and pack the resultant structure
we get a volume of $58$ (see Fig.~\ref{fig:CCZfactory}). For a fair comparison
between the distance measures of these two approaches we should rescale our
volume by a factor of $25/16$ to account for the defect diameter, so that we
obtain $91$. Such numbers tell us that braid-based surface code computation
should not be lightly dismissed in favor of lattice-surgery, despite previous
suggestions of increased storage costs \cite{Fowler12f}. Instead, a hybrid
approach to the compilation of fault-tolerant quantum circuits seems promising.

\section{Conclusions}
\label{sec:conclusions}

Although quantum computers are still far from maturity, early
proof-of-principle experiments demonstrating quantum advantage have already
been achieved \cite{Arute2019}. This sets the focus now on the pursuit of
quantum practicality for large-scale applications. To achieve this, quantum
error correction must be incorporated, adding large overheads in terms of qubit
number and time relative to preceding near-term devices. In this context it
becomes critical to reduce the size of fault-tolerant quantum circuits during
compilation.

We have presented a new approach to the problem of compressing braided quantum
circuits using the ZX-calculus as an intermediate language. The
standard interpretation of the ZX-graph identifies edges as qubits and nodes as
operation tensors. To date, considerations of cluster-state quantum computing,
even involving defects, have followed this approach \cite{Horsman11a}.  In this
work we identify that in inverting the standard interpretation we still
describe a valid quantum circuit. But in this case, the ZX-Calculus describes
braided, rather than measurement-based logic.

The use of this representation not only allows easy manipulation of the circuit
but also makes straightforward the verification of the structure of
higher-level error correction codes as well as the correctness of the
computation for Clifford sub-circuits. We have applied our compression method
to arbitrary Clifford+T circuits, including ubiquitous magic state
distilleries, reaching reduction percentages of up to  $\sim77\%$. Further, in
addition to our observed volume reductions in units of distance, the smaller
size of the circuits should reduce the required logical error rate and the
associated code distance itself.  Quantities for the code distance cannot be
estimated without a knowledge of the specific error channels and the structure
of the larger quantum computation in which the circuit is embedded. However, we
expect significant resource reductions at the physical qubit level
\cite{Litinski19a}, especially for those sub-circuits such as magic state
distilleries for which errors are heralded and the circuits can be rapidly
repeated.

It is one thing that logical information can be encoded and
manipulated in the surface code in two different ways. It is quite another to
find that the ZX-Calculus is a neutral representation of the correspondence
between them at the level of the \emph{individual logical operations}.  In the
TQC literature, the question of encoding has appeared to be settled since about
2018 \cite{Fowler18a}. Subsequent work designing efficient fault-tolerant
circuits has assumed the lattice surgery approach \cite{Litinski2019}.  Crucially, neither of our
interpretations of the ZX-Calculus for surface-based topological models
excludes the other, and thus the ZX-Calculus is a language capable of unifying
both expressions of computation in the same diagram.  With compression rates as
reported in this paper, overheads of braided, defect-based circuits are
comparable, and even lower, to those obtained for their lattice-surgery counterparts.  With the
ability to translate between these models of computation, our work not only has the
potential to resurrect braiding as a viable prospect, but to propose a new approach for computation based on hybrid circuits, an option not previously considered outside of single-qubit gate implementations. Indeed, in our own results we have found that the best
route has consisted of a hybridisation of these approaches.

Finally, prior work has focused heavily on
the reduction of a small number of common sub-circuits, for which efforts are
expected to show progressively diminishing returns. By contrast, our approach
considers larger circuits in their entirety, making reduction of the remaining
$\sim 50$\% of these circuits accessible. This work therefore opens a new
direction for the compilation of surface code based quantum computation. To generalize the applicability and scalability of our approach,
in future work we plan to explore strategies and tools to
automate the entire set of compilations and reductions.

\section{Acknowledgments}
We thank Koki Suetsugu, Kunihiro Wasa, and Yu Yokoi for interesting
discussions, as well as Austin Fowler and Craig Gidney for helpful comments on
an early preprint.
During the preparation of this manuscript, we became aware of similar work
being pursued by Alexandru Paler.
This work was supported in part by the Japanese MEXT Quantum Leap Flagship
Program (MEXT Q-LEAP), Grant Number JPMXS0118069605.

\bibliography{papers_bib}
\bibliographystyle{unsrt}

\end{document}